
\magnification=\magstep1
\define\KPZ{\operatorname{KPZ}}
\define\coS{\overline{\Cal S}}
\define\coT{\overline{\Cal T}}
\define\coB{\overline{\Cal B}}
\define\cuS{\overline{\text{\sl S}}}
\define\coG{\overline{\text{\sl G}}}
\define\cuT{\overline{\text{\sl T}}}

\define\cc{\overline{\text{\sl c}}}
\define\comu{\overline{\mu}}
\define\covar{\overline{\varphi}}
\define\cochi{\overline{\chi}}
\define\cobeta{\overline{\beta}}
\define\cgamma{\overline{\gamma}}
\define\cst.{\operatorname{cst.}}
\redefine\l{\langle}
\define\r{\rangle}
\define\rhs{\operatorname{rhs}}
\define\lhs{\operatorname{lhs}}

\TagsOnRight

\NoBlackBoxes

\topmatter
\title{\bf MULTI--SPIN SYSTEMS ON A RANDOMLY}\\
 {\bf TRIANGULATED SURFACE}
\endtitle
\author
{\rm by} B. Durhuus
\endauthor
\endtopmatter

$$
\aligned
&\text{\rm Mathematics Institute}\\
&\text{\rm University of Copenhagen}\\
&\text{\rm Universitetsparken 5}\\
&\text{\rm DK--2100 Copenhagen \O}
\endaligned
$$

\vskip12truecm

\noindent E--mail address: durhuus\@math.du.dk

\newpage

\noindent{\eightpoint{\smc abstract}.
We consider a class of spin systems on randomly
triangulated surfaces as discrete approximations to conformal matter
fields coupled to $2d$ gravity. On the basis of certain universality
assumptions we argue that at critical points with diverging string
susceptibility $(c\ge 1)$ the model either exhibits mean field
behaviour or it can effectively be described by a conformal matter
system with central charge $\cc\le 1$ coupled to $2d$ gravity. As a
particular consequence we conclude in the unitary case that the
string susceptibility exponent $\gamma$ is limited to possible values
of the form $\tfrac 1n$, $n=2,3,4,\dots\;$, where $n=2$ corresponds
to mean field behaviour.}

\vskip1truecm

\subheading{1. Introduction}
\medskip
As is well known by now two-dimensional gravity coupled to conformal
matter fields with central charge $c\le 1$ can be realized
non-perturbatively in terms of suitable spin systems on a randomly
triangulated surface. The in many respects more interesting case
$c>1$, describing string theory in physical dimensions, remains,
however, to a large extent un-understood. The problem is not a lack
of candidate models for non-perturbative formulations of string
theory in dimensions $d\ge 2$ but rather that naive formulations of
such models seem to describe point particles rather than extended
objects. E\.g\. this is the case for the naive discretization of
the Nambu--Goto string on a hypercubic lattice (see refs\. [1,2,3]).
With hindsight this is perhaps not surprising since in that case the
internal geometry of the world sheat is not allowed to fluctuate and
thus the conformal mode is abscent. However, also for the standard
naively triangulated version of the Polyakov string obtained by
embedding triangulations into $\Bbb R^d$ one finds strong
indications of instabilities leading to collaps of the world sheat.
Perhaps the strongest indication is that the string tension does
not vanish at the critical point as shown in [4].

The implications of this collaps phenomenon for the internal geometry
of the world sheat are, however, not so clear. F\.ex\. numerical
estimates of the susceptibility exponent $\gamma$ as a function of
$d$ (see ref\. [5,6]) have shown no clear sign of a discontinuity at
$d=1$ contrary to what would be expected from the known values of
$\gamma$ for $d\le 1$ given by the so-called $\KPZ$ formula
[13,14,15], if the internal geometry was that of a branched polymer
for $d>1$, in which case $\gamma$ would be $\tfrac 12$. Some
analytical investigations of this question were performed in [7].

Instead of embedding the triangulations into $\Bbb R^d$, i\.e\.
coupling two-dimensional discrete gravity to $d$ scalar fields, we
shall in this paper consider the alternative possibility of coupling
two-dimensional gravity to certain (bounded, discrete) spins with
central charge $c>1$. By combining the arguments in refs\. [1,3] with
some standard knowledge of conformal field theory we show under
certain universality
assumptions that these models are either trivial or
in a certain sense are effectively described by
models with central charge $\le 1$. This in turn implies that
$\gamma$ is constrained to the possible values
 $$
\gamma=\tfrac 1n,\quad n=2,3,4,\dots\;,
 $$
 where $n=2$ corresponds to the case where the internal geometry of
the world sheet is that of a branched polymer, whereas e\.g\. $n=3$
for a model that is effectively governed by pure gravity. In general,
at critical points for which $n>2$ the model is governed by an
effective model of central charge $1-\tfrac 6{n(n-1)}$ and the
scaling limit (to the extent it exists) will not be dominated by
tree-like surfaces alone.

As will be seen the
arguments show that this phenomenon should be expected to be
generic for a large class of spin systems with local interactions,
coupled to $2d$ gravity.

The paper is organized as follows. In section 2 we briefly recall, in
the case of spherical topology, the arguments of refs\. [1,3]. In
section 3 we introduce the models to be considered and show how those
arguments can be extended and how they lead to the constraints on
$\gamma$ quoted above. In section 4 the case of higher genus surfaces
is discussed and, in particular, it is shown that the susceptibility
exponent depends linearly on the genus. Finally, section 5 contains
some concluding remarks.

\newpage

\subheading{2. Coarse graining of surfaces and triangulations}

\medskip

For the purpose of fixing some notation and for illustrative purposes
we briefly recall in this section the arguments of refs\. [1,3]
leading to triviality of the hypercubic lattice string and show how
the basic ideas can be adapted to the triangulated version of pure
gravity on a $2$-sphere.

Let $\Cal S$ denote the set of orientable (possibly
self-intersecting or self-overlapping)
surfaces made up of plaquettes on the lattice
$\Bbb Z^d$ with the topology of a $2$-sphere with holes, i\.e\.
planar surfaces. Given links $\ell_1,\dots,\ell_n$ in $\Bbb Z^d$ let
$\Cal S_{\ell_1,\dots,\ell_n}$ denote the set of surfaces in $\Cal S$
whose boundary $\partial S$ consists of $n$ loops of length $2$, the
$i$'th loop being made up of two copies of the link $\ell_i$.
Moreover, let $\coS$ denote the subset of $\Cal S$ consisting of
surfaces containing no loops of length $2$ (except possibly boundary
loops) or, in other words, the surfaces in $\coS$ are irreducible
w\.r\.t\. cutting along loops of length $2$ (to be called $2$-loops
in the following). In analogy with $\Cal S_{\ell_1,\dots,\ell_n}$ we
let $\coS_{\ell_1,\dots,\ell_n}=\coS\cap\Cal
S_{\ell_1,\dots,\ell_n}$.

In the following we shall actually distinguish between the two copies
of a link $\ell$ in a boundary loop of a surface $S\in\Cal
S_{\ell_1,\dots,\ell_n}$ (or $\coS_{\ell_1,\dots,\ell_n})$, say, one
is white and the other black. This is for purely technical
convenience and of no fundamental importance for the arguments, and
we shall hence mako no further notice about it.

Note that $\coS$ can be regarded as being obtained by coarse graining
$\Cal S$ and that $\coS$ contains the set of planar surfaces made up
of plaquettes with sides of length $2$.

We define the $n$-point correlation functions (considering $2$-loops
as punctures) by
 $$
G_{\ell_1,\dots,\ell_n}(\mu)=\sum_{S\in\Cal S_{\ell_1,\dots,\ell_n}}
e^{-\mu|S|}\tag2.1
 $$
 and
 $$
\coG_{\ell_1,\dots,\ell_n}(\mu)=\sum_{\overline{S}
\in\coS_{\ell_1,\dots,\ell_n}}
e^{-\mu|\overline{S}|}, \tag2.2
 $$
 where $|S|$ denotes the area (number of plaquettes) in $S$ and $\mu$
is the bare string tension. Of course the one-point functions
$G(\mu)\equiv G_\ell(\mu)$ and $\coG(\mu)\equiv\coG_\ell(\mu)$ are
independent of $\ell$. It is well known ([1,3]) that
$G_{\ell_1,\dots,\ell_n}(\mu)$, resp\.
$\coG_{\ell_1,\dots,\ell_n}(\mu)$, is finite and analytic for
$\mu\ge\mu_0$, resp\. $\mu\ge\comu_0$, for some finite critical
coupling $\mu_0$, resp\. $\comu_0$, whereas they are divergent
otherwise.

We shall be interested in the scaling limits at $\mu_0$, resp\.
$\comu_0$, in the following. In view of the remarks above we expect
by universability that they coincide for the model based on  $\Cal S$
(the $\Cal S$-model), and the coarse grained one based on $\coS$ (the
$\coS$-model). In particular, their critical indices coincide.

In order to analyse the critical behaviour we consider the
susceptibilities $\chi$ and $\cochi$ given by
 $$
\chi(\mu)=-\frac{dG}{d\mu}=\sum_{S\in\Cal S_\ell}|S|e^{-\mu|S|}\tag2.3
 $$
 and
 $$
\cochi(\mu)=-\frac{d\coG}{d\mu}=
\sum_{\overline{S}\in\coS_\ell}|\cuS|e^{-\mu|\overline{S}|}.
\tag2.4
 $$
 We assume that there is a critical index $\gamma$, the string
susceptibility exponent, associated to $\chi$ given
by\footnote"*"{Here and in the following the symbol $\sim$ in a
relation indicates that the $\rhs$ represents the leading behaviour of
the $\lhs$ up to possible analytic terms that vanish as a coupling
constant approaches a critical value.}
 $$
\chi(\mu)\sim\cst.|\mu-\mu_0|^{-\gamma}\tag2.5
 $$
 as $\mu$ approaches $\mu_0$, where $\gamma>0$. In particular, we
assume that $\chi$ diverges as $\mu\to\mu_0$. Numerically this has
been checked in $d=2,3$ (see [2]). By universality, we also have
 $$
\cochi(\mu)\sim\cst.|\mu-\comu|^{-\gamma}\tag2.6
 $$
 as $\mu$ approaches $\comu_0$.

It turns out that $\gamma$ can be determined from these assumptions
alone. This is most easily accomplished by noting that the
correlation functions of the $\Cal S$-model up to inessential
boundary contributions can be expressed in terms of those of the
$\coS$-model evaluated at a certain effective coupling constant. We
shall describe this procedure for the one-- and two-point functions
which are the only cases we need.

Thus consider a surface $S\in\Cal S_\ell$. Given any 2-loop in $S$
not containing, say, the white boundary link we obtain by cutting $S$
along this loop (or along a copy of $\ell$ in case the 2-loop contains the
black link in $\partial S$) two surfaces $S'$ and $S''$ of which,
say, $S'$ contains the original white link in its boundary. It is
easy to see that any two surfaces $S_1''$ and $S_2''$ obtained by
cutting along two such 2-loops have either no link in common or
their union can also be obtained by cutting $S$ along a $2$-loop.
Thus there exists a unique collection
of maximal subsurfaces $S_i''\in\Cal S_{\ell_i}$, $i=0,1,\dots,N$
obtained in this way which pairwise have no links in common. Of
course, at most one of these, say $S_0''$, contains the black link in
$\partial S$ in its boundary, and is obtained by cutting along a copy
of $\ell$
in a 2-loop containing the black link in $\partial S$. The others are
obtained by cutting along interior 2-loops. By performing these
cuttings successively, and at each stage closing the hole (2-loop)
appearing (for $i\ne 0$) in the
$S'$-surfaces, one obtains a unique surface
$\overline{S}\in\overline{\Cal S}_\ell$ containing the
white link in $\partial S$ in its boundary. Moreover, $S$ may be
reconstructed in a unique way from $\cuS$ in an obvious way by
cutting it open along a certain set of links and gluing on the
surfaces $S_1'',\dots,S_N''$ and, in addition, by gluing $S_0''$, if
non-empty, to $\cuS$ along the black link in $\partial\cuS$. (The
reader may verify that colouring of boundary links is needed in order
to make this re-gluing process unique, say by requiring white links
to be glued to black links). We shall call $S_1'',\dots,S_N''$ {\it
outgrowths} on $S$ and $S_0''$  a {\it pocket} on
$\partial S$. See Fig\. 1.

By decomposing surfaces in $\Cal S_\ell$ in this manner we can
rewrite $G(\mu)$ as
 $$
G(\mu)=\sum_{\overline{S}\in\coS_{\ell}}e^{-\mu|\overline{S}|}
(1+G(\mu))^{2|\overline{S}|}\;,\tag2.7
 $$
 where we have summed over the outgrowths and pocket thus obtaining a
factor $1+G(\mu)$ for each interior link as well as for one boundary
link, i\.e\. $2|S|$ such factors in total. In view of (2.2) equation
(2.7) may be written as
 $$
G(\mu)=\coG(\comu)\;,\tag2.8
 $$
where the effective coupling $\comu=\comu(\mu)$ is
given by
 $$
\comu=\mu-2\log(1+G(\mu))\;.\tag2.9
 $$
 Noting that
 $$
\frac{d\comu}{d\mu}=1+\frac{2\chi(\mu)}{1+G(\mu)}\tag2.10
 $$
 we obtain
 $$
\chi(\mu)=\cochi(\comu)
\left(1-\frac{2\cochi(\comu)}{1+G(\mu)}\right)^{-1}\tag2.11
 $$
 by differentiating (2.8) w\.r\.t\. $\mu$ and solving for
$\chi(\mu)$.

{}From the positivity of $\chi$ and $\cochi$ and (2.11) we get
 $$
\cochi(\comu)<\tfrac 12(1+G(\mu))\tag2.12
 $$
 for $\mu>\mu_0$, and since $G(\mu_0)$ is finite, which is evident
from equation (2.9) since $\comu\ge\comu_0$ for $\mu>\mu_0$, it
follows that $\cochi(\comu)$ stays finite as $\mu\to\mu_0$. In view
of the fact that $\cochi$ diverges at $\comu_0$ we thus conclude that
 $$
\comu(\mu_0)\gvertneqq\comu_0\tag2.13
 $$
 and, in particular, $\coG$ and $\cochi$ are analytic at
$\comu(\mu_0)$. Moreover, since $\chi$ diverges at $\mu_0$ it follows
that
 $$
\frac{2\cochi(\comu(\mu_0))}{1+G(\mu_0)}=1\tag2.14
 $$
 and
 $$
\aligned
\chi(\mu) &\sim\cst.(\cochi(\comu(\mu_0))-\cochi(\comu))^{-1}\\
&\sim\cst.(\comu-\comu(\mu_0))^{-1}\;,
\endaligned\tag2.15
 $$
 where we used that $\cochi'(\mu)<0$ for $\mu>\comu_0$. Finally,
using that
 $$
\frac{d\comu}{d\mu}\sim\cst.(\mu-\mu_0)^{-\gamma}\tag2.16
 $$
 as $\mu\to\mu_0$ by (2.5) and (2.10) we have
 $$
\comu-\comu(\mu_0)\sim\cst.(\mu-\mu_0)^{1-\gamma}\;.\tag2.17
 $$
 From (2.5), (2.15) and (2.17) we now get
 $$
(\mu-\mu_0)^{-\gamma}\sim\cst.(\mu-\mu_0)^{\gamma-1}
 $$
 as $\mu\to\mu_0$, which yields
 $$
\gamma=\tfrac 12\;.\tag2.18
 $$

This value of $\gamma$ coincides with the mean field value obtained
in the limit $d\to\infty$ (see refs\. [1,8]). By considering in a
similar manner the one-loop function corresponding to a rectangular
boundary one may prove using (2.13) that the string tension of this
model does not scale to zero at $\mu_0$ and as a consequence that the
sum in (2.1) is dominated by tree-like surfaces characteristic of the
mean field limit. Moreover, the scaling limit describes a free
scalar field [1,9].

We note that the crucial observation in the derivation of (2.18) is
the fact (2.13) that the effective coupling $\comu$ does {\it not}
approach the critical coupling $\comu_0$ of the coarse grained model
and, in turn, this is a consequence of the divergence of the
susceptibilities.

For later purposes we shall make some remarks on the
geometric meaning of equation (2.11). For this purpose let us first
define the alternative susceptibilities $\chi_0$ and $\cochi_0$ by
 $$
\chi_0(\mu)=\sum_{\ell'}G_{\ell,\ell'}(\mu),\quad
\cochi_0(\mu)=\sum_{\ell'}\coG_{\ell,\ell'}(\mu)\;,\tag2.21
 $$
 where the sums are over all links $\ell'$ in $\Bbb Z^d$ with $\ell$
fixed. Since the number of ways to cut a surface $S\in\Cal S_\ell$
open along a link $\ell'$ so as to obtain a surface $S'\in\Cal
S_{\ell,\ell'}$ equals $2|S|$ to leading order in $|S|$ the critical
indices of $\chi_0$ and $\chi$, resp\. $\cochi_0$ and $\cochi$,
coincide. In fact, one finds that
 $$
\chi_0(\mu)=2\chi(\mu)-G(\mu)-G(\mu)^2\tag2.22
 $$
 and
 $$
\cochi_0(\mu)=2\cochi(\mu)-\coG(\mu)\;.\tag2.23
 $$

Given a fixed link $\ell'$, any surface $S\in\Cal S_{\ell,\ell'}$
contributing to $G_{\ell,\ell'}(\mu)$ can be decomposed in analogy
with the surfaces in $\Cal S_\ell$: Cutting $S$ along an interior
2-loop $L$ in $S$ consisting of two copies of a link $\ell_0$ we
obtain two surfaces $S'$ and $S''$ where, say, the boundary 2-loop in
$\partial S$ corresponding to $\ell$ is a boundary component of $S'$.
If $L$ is homotopic to the boundary 2-loop we have $S'\in\Cal
S_{\ell,\ell_0}$ and $S''\in\Cal S_{\ell_0,\ell'}$ and we say that
$L$ is transversal, see Fig\. 2. Otherwise, $L$ is contractible and
we have $S'\in\Cal S_{\ell,\ell_0,\ell'}$ and $S''\in\Cal
S_{\ell_0}$. We then call $S''$ an outgrowth of $S$, see Fig\. 2.

As in the case of surfaces in $\Cal S_\ell$ we may now first
successively eliminate maximal outgrowths as well as pockets at the
four boundary links of $S$ to obtain a unique surface $\widetilde
S\in\Cal S_{\ell,\ell'}$ all of whose 2-loops are transversal. If
there are $n$ such transversal 2-loops consisting of doublets of
links $\ell_1,\dots,\ell_n$ we finally cut $\widetilde S$ along these
to obtain $n+1$ surfaces $\cuS_1\in\Cal S_{\ell_1\ell_1}$,
$\cuS_2\in\coS_{\ell_1,\ell_2},\dots,\cuS_{n+1}\in\coS_{\ell_n,\ell'}$.

Using this decomposition procedure and noting that the total number
of factors $1+G(\mu)$ arising from the summation over outgrowths and
pockets is $2|S|+2$ we arrive at
 $$
\aligned
\chi_0(\mu) & = \sum_{n=0}^\infty
\sum_{\ell_1,\dots,\ell_n,\ell'}(1+G(\mu))^2
\coG_{\ell,\ell_1}(\comu)\coG_{\ell_1,\ell_2}(\comu)\cdot\ldots\cdot
\coG_{\ell_n,\ell'}(\comu)\\
& = (1+G(\mu))^2\sum_{n=1}^\infty\cochi_0(\comu)^n\\
& = (1+G(\mu))^2\frac{\cochi(\comu)}{1-\cochi_0(\comu)}\;,
\endaligned\tag2.24
 $$
 where $\comu$ is given by equation (2.9).

Using (2.22--23) one sees that (2.24) is equivalent to (2.11), which
may also conveniently be written as
 $$
\varphi(\mu)=\frac{\covar(\comu)}{1-\varphi(\comu)}\tag2.19
 $$
 where
 $$
\varphi(\mu)=\frac{2\chi(\mu)}{1+G(\mu)}\;,\qquad
\covar(\mu)=\frac{2\cochi(\mu)}{1+\coG(\mu)}\;.\tag2.20
 $$
 In section 3 we shall find it more convenient to generalize (2.24)
rather than (2.11).

The basic ideas of the method described above do not depend in an
essential way on the fact that surfaces are embedded in an ambient
space. As a particular illustrative example let us consider the
triangulated version of two-dimensional pure gravity [10]. Thus, let
$\Cal T$ denote the set of triangulations (abstract 2-dimensional
simplicial complexes) with the topology of a sphere with holes and
let $\Cal T_n$ denote the set of triangulations in $\Cal T$ with an
ordered set of $n$ holes (boundary components) each of which is a
2-loop, i\.e\. consists of two links. Likewise, let $\coT$ denote the
subset of $\Cal T$ consisting of triangulations without 2-loops
(except, possibly, boundary loops) and let $\coT_n=\coT\cap\Cal T_n$.

We define the $n$-point functions $G_n$ and $\coG_n$ by
 $$
G_n(\mu)=\sum_{T\in\Cal T_n}e^{-\mu|T|}\;,\quad
\coG_n(\mu)=\sum_{\overline{T}\in\coT_n}e^{-\mu|\overline{T}|}\;,\tag2.25
 $$
 where $|T|$ denotes the area, i\.e\. number of triangles in $T$, and
$\mu$ is the bare cosmological constant. There exists a finite
critical coupling constant $\mu_0$, resp\. $\comu_0$, such that
$G_n(\mu)$, resp\. $\coG_n(\mu)$, are finite analytic functions for
$\mu>\mu_0$, resp\. $\mu>\comu_0$, whereas they are infinite
otherwise, see e\.g\. refs\. [10,11,12]. Moreover, $G_n(\mu)$ equals,
to leading order in $\mu-\mu_0$, the $(n-1)$'th derivative of
$G_1(\mu)$ w\.r\.t\. $\mu$, and similarly for $\coG_n(\mu)$.

By a universality argument similar to the one applied above the two
sets of functions $G_n$ and $\coG_n$ may be considered as
representing discrete versions of one and the same continuum model
(provided it exists) and their critical behaviours hence coincide.

By arguments that are basically identical to those leading to
equation (2.8) we have
 $$
G_1(\mu)=\coG_1(\comu)\;,\tag2.26
 $$
 where the effective coupling $\comu=\comu(\mu)$ is given by
 $$
\comu=\mu-\tfrac 32\log(1+G_1(\mu))\;,\tag2.27
 $$
 the factor $\tfrac 32$ originating from the fact that the number of
links in a triangulation $T\in\Cal T_1$ is $\tfrac 32|T|+1$.
Differentiating equation (2.26) w\.r\.t\. $\mu$ yields
 $$
\chi(\mu)=\cochi(\comu)
\left(1-\frac{\frac 32\cochi(\comu)}{1+G_1(\mu)}\right)^{-1}\tag2.28
 $$
 in analogy with equation (2.11), where the susceptibilities $\chi$
and $\cochi$ are given by
 $$
\chi(\mu)=-\frac{dG_1}{d\mu}=\sum_{T\in\Cal T_1}
|T|e^{-\mu|T|}\tag2.29
 $$
 and
 $$
\cochi(\mu)=-\frac{d\coG_1}{d\mu}=\sum_{\overline{T}\in\coT_1}
|\cuT|e^{-\mu|\overline{T}|}\;.\tag2.30
 $$

Setting
 $$
\varphi(\mu)=\frac{\frac 32\chi(\mu)}{1+G_1(\mu)}\;,\quad
\covar(\mu)=\frac{\frac 32\cochi(\mu)}{1+\coG_1(\mu)}\;,\tag2.31
 $$
 equation (2.28) may be written as
 $$
\varphi(\mu)=\frac{\covar(\comu)}{1-\covar(\comu)}\;.\tag2.32
 $$
 This equation may also be derived by considering the alternative
susceptibilities
 $$
\chi_0(\mu)\equiv G_2(\mu)\;,\quad
\cochi_0(\mu)\equiv\coG_2(\mu)\tag2.33
 $$
 and decomposing the triangulations $T\in\Cal T_2$ contributing to
$G_2(\mu)$ in complete analogy with the procedure described
previously for the lattice case.

It is known for this model that the susceptibility $\chi$ does not
diverge at $\mu_0$. In fact, we have (see e\.g\. refs\. [10,12]) for
$\mu\to\mu_0$ that
 $$
\varphi(\mu)\sim\cst.(\mu-\mu_0)^{\frac 12}\tag2.34
 $$
i\.e\. the susceptibility exponent $\gamma$ is given by
 $$
\gamma=-\tfrac 12\;.\tag2.35
 $$
 Hence we cannot in this case conclude that $\comu(\mu_0)>\comu_0$.
On the contrary we have
 $$
\comu(\mu_0)=\comu_0\;.\tag2.36
 $$
 This can be seen as follows. First, differentiate equation (2.32) to
obtain
 $$
\frac{d\varphi}{d\mu}=\frac{d\covar}{d\comu}(1-\covar(\comu))^{-3}\;.\tag2.37
 $$
 Then observe that since $\varphi(\mu)$ stays finite as $\mu\to\mu_0$
it follows from (2.32) that $\varphi(\comu(\mu_0))<1$. One the other
hand $\tfrac{d\varphi}{d\mu}$ diverges as $\mu\to\mu_0$ according to
(2.34) and hence equation (2.37) implies that
$\tfrac{d\covar}{d\comu}$ also diverges as $\mu\to\mu_0$. But this is
only possible if $\comu(\mu)\to\comu_0$ as $\mu\to\mu_0$, as desired.

In addition we have
 $$
\frac{d\covar}{d\comu}\sim\cst.(\mu-\mu_0)^{-\frac 12}
\sim\cst.(\comu-\comu_0)^{-\frac 12}\;,
 $$
 since
 $$
\frac{d\comu}{d\mu}=1+\varphi(\mu)\sim\cst.-\cst.(\mu-\mu_0)^{\frac 12}
 $$
 implies
 $$
\comu(\mu)-\comu(\mu_0)\sim\cst.(\mu-\mu_0)
 $$
 for $\mu$ close to $\mu_0$. Thus the critical index of $\covar$ or
$\cochi$ is also $\gamma=-\tfrac 12$, as expected.

To sum up, we have seen that, although the method of coarse graining
can be applied to both models considered in this section, the
predictions depend crucially on whether the susceptibility diverges
or not. In the former case the coarse grained model is non-critical
and the exponent $\gamma$ may be evaluated, whereas in the latter
case the coarse grained model is critical but the value of $\gamma$
cannot be extracted from equation (2.28) alone. In the next section
we extend the analysis of the present section to a class of models
with diverging susceptibility obtained by coupling suitable spin
systems to the triangulated model of 2-dimensional gravity. These
models typically involve a set of copuling constants in addition to
the cosmological constant and we shall see that alternative
values of $\gamma$ may occur in addition to the mean field value.

\newpage

\subheading{3. Multi--Ising spins coupled to 2D gravity}
\medskip

For definiteness we shall only consider a multiplet of Ising spins on
a random triangulation, although similar considerations may be
applied to other discrete spin systems as well. Thus, for each
triangle $i$ in a triangulation $T$, let
$\sigma_i=(\sigma_i^1,\dots,\sigma_i^m)$ be an $m$'tuple whose
coordinates equal $\pm 1$. Given coupling constants $\mu$,
$\beta_1,\dots,\beta_m$ let the action associated to a spin
configuration $\sigma=\{\sigma_i\}$ be given by
 $$
A_T(\sigma)=\mu|T|-\sum_{k=1}^m \sum_{\l ij\r}
\beta_k(\sigma_i\cdot\sigma_j)^k\;,\tag3.1
 $$
where
$\sigma_i\cdot\sigma_j=\sigma_i^1\sigma_j^1+\dots+\sigma_i^m\sigma_j^m$
and the sum over $\l ij\r$ is over all pairs of nearest neighbouring
triangles, i\.e\. triangles sharing a link. Moreover, we shall
henceforth let $|T|$ denote the number of links in $|T|$. This
corresponds to rescaling $\mu$ by a factor $\tfrac 32$ compared to
the previous section.

We note that $A_T$ represents the most general nearest neighbour
interaction of the spins that is invariant under permutations of the
$m$ systems, i\.e\. under the simultaneous application of a
permutation to the coordinates of all spins $\sigma_i$, as well as
under flips of each of the systems, i\.e\. under a simulteneous
change of the $k$'th coordinate of each $\sigma_i$, for any
$k=1,\dots,m$.

The $n$-point functions $G_n$ are defined in analogy with (2.25) by
 $$
G_n(\mu,\beta)=\sum_{T\in\Cal T_n}\sum_\sigma e^{-A_T(\sigma)}\;,\tag3.2
 $$
 where the second sum is over all spin configurations on $T$. It
follows easily from general arguments in conjunction with [12] that
there exists a (non-empty) convex open domain $\Cal B\subseteq\Bbb
R^{m+1}$ such that the $G_n$ are finite analytic functions of
$(\mu,\beta)\in\Cal B$ and that they are infinit outside the closure
of $\Cal B$. On the critical surface $\partial\Cal B$ the correlation
functions display non-analytic behaviour. We shall assume that
associated to each critical point $(\mu_0,\beta_0)\in\partial\Cal B$
there is a susceptibility exponent $\gamma$ such that either the
susceptibility
 $$
\chi(\mu,\beta)\equiv G_2(\mu,\beta)\tag3.3
 $$
 diverges like
 $$
\chi(\mu,\beta)\sim\cst.|(\mu,\beta)-(\mu_0,\beta_0)|^{-\gamma}\tag3.4
 $$
 with $\gamma=\gamma(\beta_0)>0$, or it is finite at
$(\mu_0,\beta_0)$ and behaves as
 $$
\chi(\mu,\beta)\sim\cst.-\cst.|(\mu,\beta)-(\mu_0,\beta_0)|^{-\gamma}\tag3.5
 $$
 with $-1<\gamma=\gamma(\beta_0)<0$, as $(\mu,\beta)$ approaches
$(\mu_0,\beta_0)$ transversally to $\partial\Cal B$, i\.e\. along a
curve which is non-tangential to $\partial\Cal B$ (but $\gamma$ is
independent of the curve). In (3.4--5) we denote by $|\cdot|$ the
euclidean distance in $\Bbb R^{m+1}$.

In fact, if the susceptibility is finite at $(\mu_0,\beta_0)$ it
seems reasonable to assume that there is an associated scaling limit
describing conformal matter with $c<1$ coupled to 2D gravity. After
all we do not know any examples of such systems with $c>1$ and
$\gamma<0$. In addition the scaling limit would be expected to be
unitary since the action (3.1) is reflection positive on a regular
lattice. As is well known [13,14,15] this limits the possible values
of $\gamma$ at such points to $-\tfrac 1k$, $k=2,3,4,\dots\;$. We
shall return to this assumption later in this section.

As in the previous section we define a $\coT$-model whose $n$-point
functions $\coG_n$ are given by (3.2) with $\coT_n$ replacing $\Cal
T_n$, and which are finite and analytic in a convex open region
$\coB$. The main point to observe is that universality is reflected
in the equivalence of the phase structure of the two models, i\.e\.
there is a bijective correspondence between critical points on
$\partial\Cal B$ and $\partial\coB$ with identical scaling behaviour.

In order to carry through the coarse graining procedure of the
previous section we shall have to introduce correlation functions
subject to suitable boundary conditions. It suffices for us to
consider the one-- and two-point functions. Thus, let
 $$
\widetilde H_{\sigma_1,\sigma_2}(\mu,\beta)=
\sum_{T\in\Cal T_1}\sum_{\sigma|\partial T=(\sigma_1,\sigma_2)}
e^{-A_T(\sigma)}\;,\tag3.6
 $$
 where the second sum is over all spin
configurations $\sigma$ whose two boundary spins are fixed to
$\sigma_1$ and $\sigma_2$, respectively. Here, of course, the
boundary spins are those located in the boundary triangles, i\.e\.
the triangles containing a boundary link. Moreover, let
 $$
H_{\sigma_1,\sigma_2}(\mu,\beta)=\sum_{\sigma_1',\sigma_2'}
e^{\tsize{\sum\limits_{k=1}^m}\beta_k((\sigma_1\cdot\sigma_1')^k
+(\sigma_2\cdot\sigma_2')^k)}H_{\sigma_1',\sigma_2'}(\mu,\beta)\;,\tag3.7
 $$
which can be regarded as the one-point function obtained by coupling
the boundary spins $\sigma_1',\sigma_2'$ to two fixed spins
$\sigma_1,\sigma_2$ outside the boundary.

By the symmetries of the action $A_T(\sigma)$ it follows that
$H_{\sigma_1,\sigma_2}(\mu,\beta)$ as well as $\widetilde
H_{\sigma_1,\sigma_2}(\mu,\beta)$ only depends on
$s=\sigma_1\cdot\sigma_2$. Hence, we shall write
$H_{\sigma_1\cdot\sigma_2}(\mu,\beta)$ for
$H_{\sigma_1,\sigma_2}(\mu,\beta)$. Since the possible values of
$s=\sigma_1\cdot\sigma_2$ are $-n,-n+2,\dots,n-1,n$, there are in
total $n+1$ different one-point functions $H_s(\mu,\beta)$.
Obviously, the critical indices of these are identical to that of
$G_1(\mu,\beta)$.

The relevance of the one-point functions $H_s(\mu,\beta)$ in the
coarse graining process follows by noting that summation over
outgrowths emerging from an interior link $\ell$ in a triangulation
$\widetilde T$ yields an effective contribution associated with the
pair $\l ij\r$ of triangles in $\widetilde T$ sharing $\ell$, that
depends on the spins $\sigma_i$ and $\sigma_j$ on those triangles and
is given by
 $$
e^{-\mu+\tsize{\sum\limits_{k=1}^m}\beta_k(\sigma_i\cdot\sigma_j)^k}+
H_{\sigma_i\cdot\sigma_j}(\mu,\beta)\;,\tag3.8
 $$
 where the first term corresponds to the empty outgrowth. This
contribution may be accounted for by an action associated with $\l
ij\r$ of the same form as in (3.1) with effective couplings
$(\comu,\cobeta)$ fulfilling
 $$
e^{-\comu+\tsize{\sum\limits_{k=1}^m}\cobeta_ks^k}=
e^{-\mu+\tsize{\sum\limits_{k=1}^m}\beta_ks^k}+
H_s(\mu,\beta)\tag3.9
 $$
 for $s=-n,-n+2,\dots,n-2,n$. Indeed, taking logarithms this yields
the linear system of $n+1$ equations
 $$
(\mu-\comu)+\sum_{k=1}^m(\cobeta_k-\beta)s^k=
\log\left(1+e^{\mu-\tsize{\sum\limits_{k=1}^m}\beta_ks^k}H_s(\mu,\beta)\right)\;,\tag3.10
 $$
which has a unique solution
$(\comu,\cobeta)=(\comu(\mu,\beta),\cobeta(\mu,\beta))$, the
determinant of the coefficient matrix being the van der Monde
determinant over $-n,-n+2,\dots,n-2,n$. Thus, setting
 $$
\widetilde G_s(\mu,\beta)=
e^{\mu-\tsize{\sum\limits_{k=1}^m}\beta_ks^k}H_s(\mu,\beta)\;,\tag3.11
 $$
 there exist constants $a_s$ and $b_{ks}$ such that
 $$
\align
\comu & = \mu-\sum_s a_s\log(1+\widetilde G_s(\mu,\beta))\tag3.12\\
\cobeta_k & = \beta_k+\sum_s b_{ks}\log(1+\widetilde
G_s(\mu,\beta))\;,\tag3.13
\endalign
 $$
 generalizing (2.9) and (2.27).

We remark that even for a standard Ising coupling, i\.e\. for
$\beta_2=\dots=\beta_m=0$, we generally have that $\cobeta_k\ne 0$
for $k=2,\dots,m$ (and $m\ge 2$). This is, of course, the reason for
introducing the general action (3.1).

On the basis of these observations we may now adopt the arguments of
the previous section in a straightforward way. Thus for the one-point
function we obtain, up to boundary corrections,
 $$
G_1(\mu,\beta)=\coG_1(\comu,\cobeta)\tag3.14
 $$
 in analogy with (2.26) with $(\comu,\cobeta)$ given by (3.12--13).
To be precise, equation (3.14) holds if $G_1$, resp\. $\coG_1$, is
defined such that no factors $e^{-\mu}$, resp\. $e^{-\comu}$, are
associated to boundary links, and in addition no pockets are allowed at the
boundary links of the triangulations $T$ contributing to $G_1$,
i\.e\. neither of the two boundary links is contained in any 2-loop
in $T$ except the boundary loop (this is, of course, automatic for
$\coG_1$).

The analogue of equation (2.28) is most conveniently obtained in
matrix form by imposing suitable boundary conditions on the
triangulations $T$ and spin configurations $\sigma$ contributing to
$G_2(\mu,\beta)$ in (3.2). First, $T$ is restricted to have no
pockets at the boundary links. Second, the spin configurations are
restricted such that the values of the two pairs of boundary spins at
the two ends (or boundary 2-loops) of $T$ are fixed, say, to
$(\sigma_1,\sigma_2)$, $(\sigma_1',\sigma_2')$. Finally, no factor
$e^{-\mu}$ is associated to the boundary links. In this way we obtain
for each choice of $((\sigma_1,\sigma_2),(\sigma_1',\sigma_2'))$ a
two-point function, or susceptibility, and these can be collected
into a susceptibility matrix $\chi_0(\mu,\beta)$ whose rows and
columns are labelled by $(\sigma_1,\sigma_2)$ and
$(\sigma_1',\sigma_2')$, respectively. Similarly, a susceptibility
matrix $\cochi_0(\mu,\beta)$ is defined for the coarse grained model
(for which, however, the constraint on the triangulations is empty).
Clearly, the critical behaviour of all matrix elements of
$\chi_0(\mu,\beta)$ is governed by the same critical exponent
$\gamma$ as for $\chi(\mu,\beta)$, i\.e\. they either all diverge at
the same rate given by (3.4) at a critical point $(\mu_0,\beta_0)$ or
they are all finite and behave as in (3.5).

The decomposition procedure of the previous section for
triangulations $T\in\Cal T_2$ is now seen to yield
 $$
\chi_0(\mu,\beta)=\cochi_0(\comu,\cobeta)
(1-K\cochi_0(\comu,\cobeta))^{-1}\;,\tag3.15
 $$
 where the matrix $K=K(\comu,\cobeta)$ has matrix elements given by
 $$
K_{(\sigma_1,\sigma_2),(\sigma_1',\sigma_2')}=
e^{-2\comu+\tsize{\sum\limits_{k=1}^m}((\sigma_1\cdot\sigma_1')^k+(\sigma_2\cdot\sigma_2')^k)}\tag3.16
 $$
 and represents the coupling across a transversal 2-loop, see Fig\. 2.

Let us now consider a critical point $(\mu_0,\beta_0)\in\partial\Cal
B$ at which the susceptibility diverges. It is natural to expect the
existence of such points for $m>2$ since, in particular, couplings of
two-dimensional gravity to matter fields with central charge
$c=\tfrac m2>1$ are represented by points $(\mu,\beta)$ with
$\beta_2=\dots=\beta_m=0$ and since $\gamma$ is an increasing
function of $c$, for $c<1$, with $\gamma=0$ for $c=1$ according to
the well-known relation between $\gamma$ and $c$ for $c\le 1$
[13--15]. Indeed, numerical estimates [16,17] have confirmed this
espectation. For generic points on $\partial\Cal B$ we expect,
however, that the spin degrees of freedom do not couple effectively
to the surface geometry and thus represent pure gravity with
convergent susceptibility and $\gamma=-\tfrac 12$, a phenomenon first
exhibited in [18], see also [19]. As discussed previously in this
section we can more generally not exclude critical points with
convergent susceptibility corresponding to unitary matter with $c<1$
coupled to two-dimensional gravity for which $\gamma=-\tfrac 1k$,
$k=2,3,4,\dots\;$.

We first observe that equation (3.15) in conjunction with positivity
of the matrix elements of $\chi_0(\mu,\beta)$ and
$\cochi_0(\comu,\cobeta)$ immediately implies that
$\cochi_0(\comu,\cobeta)$ ({\it or}
$\cochi(\comu,\cobeta))$ {\it does not diverge as}
$(\mu,\beta)\to(\mu_0,\beta_0)$. There are thus two possibilities:
\roster
 \item"1)" $(\comu,\cobeta)$ does not approach the critical surface
as $(\mu,\beta)\to(\mu_0,\beta_0)$, in which case $\cochi_0$ is
analytic at $(\comu(\mu_0,\beta_0),\cobeta(\mu_0,\beta_0))$,
 \item"2)" $(\comu,\cobeta)$ approaches a critical point
$(\comu_0,\cobeta_0)=(\comu(\mu_0,\beta_0),
\cobeta(\mu_0,\beta_0)) \hfil\break \in\partial\Cal B$ at
which the susceptibility $\cochi$ is finite
with susceptibility exponent $\cgamma\le 0$.
 \endroster

For simplicity, let $(\mu,\beta)$ approach $(\mu_0,\beta_0)$ along a
line parallel to the $\mu$-axis, i\.e\. let $\beta=\beta_0$ and
$\mu\to\mu_0$. From equation (3.12) we have (assuming no accidental
cancellations)
 $$
\frac{\partial\comu}{\partial\mu}\sim\cst.
\chi(\mu,\beta_0)\sim\cst.(\mu-\mu_0)^{-\gamma}\tag3.17
 $$
 and hence
 $$
\comu-\comu(\mu_0,\beta_0)\sim\cst.(\mu-\mu_0)^{1-\gamma}\;.\tag3.18
 $$
Similarly by (3.13)
 $$
\cobeta_k-\cobeta_k(\mu_0,\beta_0)\sim\cst.(\mu-\mu_0)^{1-\gamma}\;.\tag3.19
 $$
 In case 1) we have
 $$
\align
K\cochi_0(\comu,\cobeta) & \sim A-B|(\comu,\cobeta)-
(\comu(\mu_0,\beta_0),\cobeta(\mu_0,\beta_0)|\\
& \sim A-B'(\mu-\mu_0)^{1-\gamma}\;,\tag3.20
\endalign
 $$
 whereas in case 2) we have
 $$
\align
K\cochi_0(\comu,\cobeta) & \sim A-B|(\comu,\cobeta)-
(\comu_0,\cobeta_0)|^{-\cgamma}\\
& \sim A-B'(\mu-\mu_0)^{(\gamma-1)\cgamma}\;,\tag3.21
\endalign
 $$
 where $A$, $B$, $B'$ are constant matrices, and we have assumed, in
the latter case, that $(\comu,\cobeta)$ approaches $\partial\coB$
transversally.

Rewriting equation (3.15) as
 $$
\chi_0(\mu,\beta)(1-K\cochi_0(\comu,\cobeta))=\cochi_0(\comu,\cobeta)\tag3.22
 $$
 and setting
 $$
\chi_0(\mu,\beta_0)\sim M(\mu-\mu_0)^{-\gamma}\tag3.23
 $$
 where $M$ is a constant matrix, we get by inserting (3.20) and
(3.23) into (3.22) that
 $$
M(\mu-\mu_0)^{-\gamma}(1-A+B'(\mu-\mu_0)^{1-\gamma})\sim C\;,\tag3.24
 $$
 where $C=\cochi_0(\comu(\mu_0,\beta_0),\cobeta(\mu_0,\beta_0))$. The
relation (3.24) yields $M(1-A)=0$ and
 $$
1-2\gamma=0\;,\qquad\text{i\.e\.}\quad
\gamma=\tfrac 12\;.\tag3.25
 $$
 Similarly, inserting (3.21) and (3.23) into (3.22) we get $M(1-A)=0$
and
 $$
-\gamma+(\gamma-1)\cgamma=0
 $$
 i\.e\.
 $$
\gamma=\frac{\cgamma}{\cgamma-1}\;.\tag3.26
 $$

We have here discarded the subleading (non-universal) terms in
$\chi(\mu,\beta)$. This is justified if $M$ happens to be regular, in
which case $A=1$. An alternative derivation of (3.26) independent of
this detail can be based on the scalar equation obtained by taking
the trace of (3.15), in which the divergence of $\chi$ is caused by
an eigenvalue of $K\cochi_0(\comu,\cobeta)$ converging to $1$, the
convergence being governed by the exponent $\cgamma$.

Inserting $\cgamma=-\tfrac 1k$ into (3.26) one gets $\gamma=\tfrac
1{k+1}$, $k=2,3,\dots\;$. Thus equations (3.25--26) together yield
 $$
\gamma=\tfrac 1{k+1}\;,\qquad
k=1,2,3,4,\dots\;,\tag3.27
 $$
 where $k=1$ corresponds to case 1) above, in which treelike surfaces
dominate (see also next section), whereas for $k=2,3,\dots$ the
model is effectively geverned by a model with central charge
 $$
\cc=1-\frac 6{k(k+1)}\;,
 $$
 coupled to two-dimensional gravity.

We are, unfortunately, not able to determine the dependence of the
effective central charge $\cc$ on the number $m$ of Ising copies
(nor on $\beta_0$). In [19] a limit for $m\to\infty$ with
$\beta_2=\dots=\beta_m=0$ was constructed and shown to exhibit a
transition between a pure gravity phase with $\gamma=-\tfrac 12$ and
a branched polymer phase with $\gamma=\tfrac 12$, while at the
transition point an intermediate phase with $\gamma=\tfrac 13$ was
found. The model considered in [19] is basically equivalent to the
matrix models considered in [20,21], see also [22] for related work.
Whether intermediate values of $k$ between $2$ (for $m\to\infty$) and
$\infty$ (for $m=2$) occur is at present not known and seems to be
difficult to ascertain numerically.

{}From the general nature of the arguments presented we expect them to
apply to a wide class of (not necessarily unitary) models with
local interactions exhibiting second order phase transitions in flat
space, although the details will be different and perhaps complicated
to work out. In such cases an equation of the form (3.15) will hold
and one can conclude that either $\gamma=\tfrac 12$ or $\gamma$ is
given by equation (3.16) where $\cgamma$ is the susceptibility
exponent associated with a $\cc<1$ conformal model coupled to $2d$
gravity given by [13--15]
 $$
\cgamma=\frac{\cc-1-\sqrt{(1-\cc)(25-\cc)}}{12}\;.\tag3.28
 $$
Inserted into (3.16) this yields
 $$
\gamma=\frac{\cc-1+\sqrt{(1-\cc)(25-\cc)}}{12}\;.\tag3.29
 $$
 Needless to say the determination of $\cc$ remains a crucial issue.

\newpage

\subheading{4. Higher genus triangulations}

\medskip

We shall in this section briefly discuss the extension to spin
systems on higher genus triangulations of the arguments presented in
the previous sections.

By $\Cal T_n^g$ we denote the set of trianglations of a surface with
$g$ handles and $n$ punctures, i\.e\. with boundary consisting of $n$
2-loops, and we set
 $$
G_n^g(\mu,\beta)=\sum_{T\in\Cal T_n^g}\sum_{\sigma}
e^{-A_T(\sigma)}\;,\tag4.1
 $$
 where $A_T(\sigma)$ is given by (3.1). Thus
$G_n^0(\mu,\beta)=G_n(\mu,\beta)$. It can be shown that the functions
$G_n^g$ are all finite and analytic in the region $\Cal B$ but
divergent outside the closure of $\Cal B$. Similarly, one defines
$\coG_n^g$ by replacing $\Cal T_n^g$ by $\coT_n^g$ in (4.1), where
$\coT_n^g$ consists of triangulations in $\Cal T_n^g$ without 2-loops
(except for the boundary loops), and which are analytic in the region
$\coB$ but divergent outside the closure of $\coB$.

The susceptibility exponents $\gamma^g=\gamma^g(\beta)$ and
$\cgamma^g=\cgamma^g(\beta)$ are defined by
 $$
G_2^g(\mu,\beta)\sim\cst.|\mu-\mu_0(\beta)|^{-\gamma^g}
 $$
 and
 $$
\coG_2^g(\mu,\beta)\sim\cst.|\mu-\comu_0(\beta)|^{-\cgamma^g}
 $$
 as $\mu\to\mu_0(\beta)$ and $\mu\to\comu_0(\beta)$, respectively,
anticipating that they are $>0$ for $g\ge 1$. Note that since the
$n$-point function essentially equals the $(n-1)$'th derivative of
the $1$-point function w\.r\.t\. $\mu$ we have
 $$
G_n^g(\mu,\beta)\sim\cst.|\mu-\mu_0(\beta)|^{2-\gamma^g-n}\tag4.2
 $$
 and similarly for $\coG_n^g$.

In carrying out the coarse graining process of the previous sections
no basic additional complications arise except that the number of
possible ways of cutting surfaces proliferates as $g$ increases. We
shall confine ourselves to considering the cases $g=1,2$, from which
the general picture will be obvious.

Given an interior $2$-loop $L$ in a triangulation $T\in\Cal T_1^g$
three cases may occur:
\roster
 \item"a)" $L$ is homotopic to a point, i\.e\. contractible in $T$.
 \item"b)" $L$ is homotopic to the boundary loop.
 \item"c)" $L$ is neither contractible nor homotopic to the boundary.
 \endroster
In case a) the loop $L$ bounds an outgrowth on $T$
and any two such outgrowths are either disjoint or their union is
also an outgrowth. By summing over maximal outgrowths with spin configurations
first, we can according to the previous section rewrite $G_1^g$ (up to
boundary corrections) as
 $$
G_1^g(\mu,\beta)=\sum_{\widetilde T}\sum_{\sigma}
e^{-\comu|\widetilde T|+\sum_{k=1}^m\sum_{\l
ij\r}\cobeta_k(\sigma_i\cdot\sigma_j)^k}\;,\tag4.3
 $$
 where the sum over $\widetilde T$ runs over triangulations in $\Cal
T_1^g$ containing no contractible $2$-loops and where $\comu$ and
$\cobeta$ are given by equations (3.12--13).

Cutting $\widetilde T$ along a $2$-loop homotopic to the boundary
splits $\widetilde T$ into a cylindrical triangulation $\widetilde
T_1\in\Cal T_2$ containing $\partial\widetilde T$ and a triangulation
$\widetilde T_2\in\Cal T_1^g$. Clearly, there is a unique maximal
such cylindrical triangulation for which $\widetilde T_2$ has no
$2$-loops homotopic to the boundary. Summation over the maximal
cylindrical triangulations $\widetilde T_1$ yields (cf\. the
decomposition of the $2$-point function in the previous section) a
factor of order $\chi(\mu,\beta)$, i\.e\. we have
 $$
G_1^g(\mu,\beta)\sim\chi(\mu,\beta)\sum_{\widetilde T}
\sum_{\sigma}e^{-\comu|\widetilde T_2|+\sum_{k=1}^m\sum_{l
ij\r}\cobeta_k(\sigma_i\cdot\sigma_j)^k}\;,\tag4.4
 $$
 where the sum over $\widetilde T_2$ is over triangulations in $\Cal
T_1^g$ all of whose $2$-loops are non-contractible and not homotopic
to the boundary $2$-loop.

In the case $g=1$ we may split the sum over $\widetilde T_2$ in (4.4)
in two parts depending on whether $\widetilde T_2$ contains any
$2$-loops (different from the boundary) or not, i\.e\.
whether $\widetilde T_2\notin\coT_1^g$ or $\widetilde
T_1\in\coT_2^g$. The latter sum yields simply
$\coG_1^1(\comu,\cobeta)$. On the other hand, a $\widetilde T_2$
containing at least one $2$-loop contains a unique maximal
cylindrical subtriangulation $\widetilde{T}_2'$ bounded by two
$2$-loops (possibly the degenerate one consisting of a single
$2$-loop). Cutting $\widetilde T_2$ along these $2$-loops splits
$\widetilde T_2$ into $\widetilde T_2'$ and a triangulation
$\widetilde T_2''\in\coT_3$. Thus summation in (4.4) over $\widetilde
T_2\notin\coT_1'$ is equivalent to summing over $\widetilde T_2'$ and
$\widetilde T_2''$ independently and yields a contribution of order
$\chi(\mu,\beta)\coG_3(\comu,\cobeta)$. In total we have
 $$
G_1^1(\mu,\beta)\sim\chi(\mu,\beta)\coG_1^1(\comu,\cobeta)+
(\chi(\mu,\beta))^2\coG_3(\comu,\cobeta)\tag4.5
 $$
 (where constant factors on the $\rhs$ have been dropped). The two
terms on the $\rhs$ and the associated decomposition of
triangulations are illustrated on Fig\. 3 where solid lines represent
the susceptibility (propagator) $G_2(\mu,\beta)$, whereas a shaded
blob with $g$ holes and $n$ dots on the outer boundary represents
$\coG_n^g(\comu,\cobeta)$.

Similarly, one finds 12 terms for $g=2$
 $$
\align
G_1^2(\mu,\beta) & \sim\chi(\mu,\beta)\coG_1^2(\comu,\cobeta)+
(\chi(\mu,\beta))^2\coG_1^1(\comu,\cobeta)\coG_2^1(\comu,\cobeta)\\
& +\dots +(\chi(\mu,\beta))^5(\coG_3(\mu,\beta))^3\;.\tag4.6
\endalign
$$
The three terms written explicitly on the $\rhs$ of (4.6) are
depicted in Fig\. 4.

{}From (4.5--6) we can now calculate $\gamma^1$ and $\gamma^2$ assuming
the scenario presented in the previous section. In case 1) where
$(\comu(\mu_0,\beta_0)$, $\cobeta(\mu_0,\beta_0))$ is not a critical
point on $\partial\coB$ the $\coG_n^g(\comu,\cobeta)$ are finite as
$\mu\to\mu_0(\beta_0)$. Since $\chi(\mu,\beta)$ diverges the leading
terms on the $\rhs$ of (4.5--6) are those containing the highest
power of $\chi(\mu,\beta)$, i\.e\. the largest number of propagators.
This, of course, is a reflection of the tree-like structure of the
dominant triangulations. Thus we get
 $$
G_1^1(\mu,\beta)\sim\cst.(\chi(\mu,\beta))^2\sim\cst.(\mu-\mu_0)^{-1}
 $$
 and
 $$
G_1^2(\mu,\beta)\sim\cst.(\chi(\mu,\beta))^5\sim\cst.(\mu-\mu_0)
^{-\frac 52}\;,
 $$
since $\gamma^{\cdot}=\tfrac 12$ in this case. More generally it is
easy to see that for arbitrary $g\ge 1$ the largest number of
propagators occurring in the decomposition process is $3g-1$ and one
finds that
 $$
G_1^g(\mu,\beta)\sim\cst.(\mu-\mu_0)^{\frac 12-\frac 32 g}\tag4.7
 $$
 which by (4.2) means that $\gamma=\tfrac 12+\tfrac 32 g$ or
 $$
\gamma^g=\gamma+g(2-\gamma)\tag4.8
 $$
 which is a well known formula for $c<1$ [13,14,15].

On the other hand, in case 2) where $(\comu(\mu_0,\beta_0)$,
$\cobeta(\mu_0,\beta_0))$ equals a critical point
$(\comu_0,\cobeta_0)\in\partial\coB$ we have
 $$
\align
\coG_n^g(\comu,\cobeta) & \sim\cst.|(\comu,\cobeta)-
(\comu_0,\cobeta_0)|^{2-\cgamma^g-n}\\
& \sim\cst.(\mu-\mu_0)^{(1-\gamma)(2-\cgamma^g-n)}\;,\tag4.9
\endalign
 $$
 where we have used the definition of $\cgamma^g$ and (3.18).
Moreover, by the analogue of (4.8) for $c<1$ we have
 $$
\cgamma^g=\cgamma+g(2-\cgamma)\;.\tag4.10
 $$
 Inserting (4.9) into the $\rhs$ of (4.5) or (4.6) and using (3.26) and
(4.10) one finds that {\it all terms are of the same order} as
$\mu\to\mu_0$ and that equation (4.8) holds in this case as well.
That this is true also for $g>2$ is not difficult to see.

Finally, we point out that for the case $c<1$, e\.g\. for a single
Ising spin, the same argument as given in section 2 shows that
$(\comu(\mu_0,\beta_0),\cobeta(\mu_0,\beta_0))$ is a critical point
on $\partial\coB$. However, in this case the susceptibility
$\chi(\mu,\beta)$ is finite at $(\mu_0,\beta_0)$ and it follows that
the term of the $\rhs$ of (4.5--6) with the fewest (one) propagators
is dominant. Evidently, this yields
 $$
\gamma^g=\cgamma^g\;,
 $$
 which expresses universality in coarse graining for arbitrary $g$ in
case $c<1$ contrary to the case $c>1$ where universality is realized
in a more subtle way, if realized at all.

\newpage
\subheading{5. Discussion}

\medskip

We have in this paper presented arguments for a scenario according to
which a class of spin systems on a random triangulation with
diverging susceptibility is either trivial or after coarse graining
may effectively be described by a $\cc\le 1$ model coupled to $2d$
gravity. In the unitary case this leads to the constraint (3.27) on
the possible values of the string susceptibility exponent, which for
$k\ge 2$ corresponds to a scaling limit with non-mean field
behaviour. Our arguments rely heavily on the assumption that scaling
limits at critical points with finite susceptibility describe
conformal matter fields with central charge $\cc\le 1$ coupled to
$2d$ gravity. It is obviously of interest to verify the existence of
such points and more generally to investigate the phase diagram for
the class of models considered here, possibly numerically.

As mentioned in the text the generic value of $\cc$ is expected to be
zero implying that $\gamma$ generically would be expected to be
$\tfrac 13$, provided the model is non-trivial. For the multi-Ising
models considered in this paper the value $\gamma=\tfrac 13$ is
realized at certain critical points in the limit where the number of
Ising copies is infinite [20]. As discussed in section 2 the value
 $\gamma=\tfrac 12$ is realized for the
hypercubic lattice string. In refs\. [24,25] a
restriction of this model was considered in which surfaces with
backtracking plaquettes are excluded. Numerical estimates
surprisingly gave a value of $\gamma$
equal to $\tfrac 14$ within errors of a few
percent in four dimensions. Coarse graining arguments for a more
general class of models with three coupling constants containing both
of these models (actually the latter as a limiting case) were carried
out in [26], and it was concluded on the basis of the observed value
of $\gamma=\tfrac 14$ that the effective model in dimension four
would be critical with finite susceptibility. In view of the
arguments of section 3 we can even conclude that the susceptibility
exponent of the effective model is close to $-\tfrac 13$, suggesting
on the basis of unitarity that this critical point corresponds to a
$c=\tfrac 12$ conformal field coupled to $2d$ gravity. To our
knowledge this would be the first realization of a $c<1$ model
coupled to $2d$ gravity in terms of embedded surfaces. Evidently, a
search for such a critical point, possibly numerically, would be of
interest.

\newpage

\noindent{\smc acknowledgements}. The author acknowledges the kind
hospitality extended to him at the Yukawa Institute for Theoretical
Physics, Kyoto, in the fall of 1992, where most of this work was
done. Likewise valuable discussions with J. Ambj\o rn and
T. J\'onsson on related subjects are acknowledged.

\vskip2truecm

\Refs\nofrills{References}

\ref[1]\by B. Durhuus, J. Fr\"ohlich, T. J\'onsson\jour Nucl.
Phys. {\bf B 240} [FS 12] (1984) p\. 453\endref

\ref[2]\by H. Kawai\jour Okamoto, Phys. Lett. {\bf 130 B} (1983), p\.
415\endref

\ref[3]\by B. Durhuus\inbook Constructive quantum field theory
II\publ eds\. G. Velo, A. S. Wightman\publaddr Plenum Press, 1990\endref

\ref[4]\by J. Ambj\o rn, B. Durhuus\jour Phys. Lett. {\bf 188 b}
(1987) p\. 253\endref

\ref[5]\by F. David\jour Nucl. Phys. {\bf B 257} (1985), p\.
543\endref

\ref[6]\by J. Ambj\o rn, B. Durhuus, J. Fr\"ohlich, P. Orland\jour
Nucl. Phys. {\bf 270} (1986), p\. 457\endref

\ref[7]\by J. Fr\"ohlich\inbook Recent developments in quantum field
theory\publ eds\. J. Ambj\o rn, B. Durhuus, J. L. Petersen\publaddr
North-Holland 1985\endref

\ref[8]\by J. M. Drouffe, G. Parisi, N. Sourlas\jour Nucl. Phys. {\bf
B 161} (1980), p\. 397\endref

\ref[9]\by T. J\'onsson\jour Comm. Math. Phys. {\bf 106} (1986), p\.
679\endref

\ref[10]\by F. David\jour Nucl. Phys. {\bf B 257} (1985), p\.
45\endref

\ref[11]\by J. Ambj\o rn, B. Durhuus, J. Fr\"ohlich\jour Nucl. Phys.
{\bf B 257} (1985), p\. 433\endref

\ref[12]\by W. T. Tutte\jour Can. Journ. Math. {\bf 14} (1962), p\.
21\endref

\ref[13]\by V. G. Kniznik, A. M. Polyakov, A. B. Zamolodchikov\jour
Mod. Phys. Lett. {\bf A3} (1988), p\. 819\endref

\ref[14]\by F. David\jour Mod. Phys. Lett. {\bf A3} (1988), p\.
1651\endref

\ref[15]\by J. Distler, H. Kawai\jour Nucl. Phys. {\bf B 321} (1989),
p\. 509\endref

\ref[16]\by E. Brezin, S. Hikami\jour Phys. Lett. {\bf B 295} (1992),
p\. 209\endref

\ref[17]\by S. Hikami\jour Phys. Lett. {\bf B 305} (1993), p\.
327\endref

\ref[18]\by V. Kazakov\jour Phys. Lett. {\bf 119 B} (1986), p\.
140\endref

\ref[19]\by D. Boulatov, V. Kazakov\jour Phys. Lett. {\bf B 186}
(1986), p\. 379\endref

\ref[20]\by M. Wexler\jour Phys. Lett. {\bf B 315} (1993), p\. 67 and Mod.
Phys. Lett. {\bf A8} (1993), p\. 2703\endref

\ref[21]\by G. P. Korchemsky\jour Phys. Lett. {\bf B 296} (1992), p\.
323 \moreref ``Matrix models perturbed by higher curvature terms'',
UPRF--92--339, preprint\endref

\ref[22]\by S. R. Das, A. Dhar, A. M. Sengupta, S. R. Wadia\jour Mod.
Phys. Lett. {\bf A5} (1990), p\. 1041\endref

\ref[23]\by L. Alvarez-Gaum\'e, J. L. F. Barbon, C. Crnkovic\jour Nucl.
Phys. {\bf B 394} (1993), p\. 383\endref

\newpage

\centerline{\smc figure captions}

\vskip1truecm

\roster
 \item"Fig\. 1."  Decomposition of a surface $S\in\Cal S_\ell$. The
subsurface $O$ is an outgrowth and $P$ is a pocket. $W$ (white) and $B$
(black) indicate the two boundary links.
 \item"Fig\. 2." Decomposition of a surface $S\in \Cal
S_{\ell,\ell'}$. The transversal $2$-loops are denoted by
$\ell_1,\ell_2$, and $O$ indicates an outgrowth.
 \item"Fig\. 3." Graphs representing the two terms on the $\rhs$ of
eq\. (4.5) and the associated decomposition of a triangulation in
$\Cal T_1^1$.
 \item"Fig\. 4." Graphs representing the three terms on the $\rhs$ of
eq\. (4.6) and the associated decomposition of a triangulation in
$\Cal T_1^2$.
 \endroster

\enddocument